\documentclass[aip,jcp,reprint,superscriptaddress,citeautoscript]{revtex4-1}
\usepackage[latin1]{inputenc}
\usepackage{graphicx}
\usepackage{fixmath}
\usepackage[colorlinks=true,linkcolor=black,%
citecolor=black,urlcolor=black]{hyperref}

\usepackage{amssymb}
\usepackage{amsmath}
\usepackage{amsthm}
\usepackage{bm}
\usepackage{multirow}
\usepackage{psfrag}
\usepackage{mathrsfs}
\usepackage[normalem]{ulem} 
\usepackage{tabularx}

\newcommand{\etal}{\emph{et al.}}

\DeclareMathOperator{\erfc}{erfc}
\DeclareMathOperator{\erf}{erf}

\newcommand{\imwf}{Institut f\"ur Materialpr\"ufung, Werkstoffkunde und Festigkeitslehre (IMWF),
  Universit\"at Stuttgart, Pfaffenwaldring 32, 70550 Stuttgart, Germany}
\newcommand{\itap}{Institut f\"ur Theoretische und Angewandte Physik (ITAP),
  Universit\"at Stuttgart, Pfaffenwaldring 57, 70550 Stuttgart, Germany}

\begin{document}
\title{Simulation of crack propagation in alumina with ab-initio based polarizable force field}

\date{\today}

\author{Stephen Hocker}
\affiliation{\imwf}
\author{Philipp Beck}
\affiliation{\itap}
\author{Siegfried Schmauder}
\affiliation{\imwf}
\author{Johannes Roth}
\affiliation{\itap}
\author{Hans-Rainer Trebin}
\affiliation{\itap}

\begin{abstract}
We present an effective atomic interaction potential for crystalline $\alpha$-Al$_{2}$O$_{3}$ generated by the program \emph{potfit}. 
The Wolf direct, pairwise summation method with 
spherical truncation is used for electrostatic interactions. 
The polarizability of oxygen atoms is included by use of the Tangney-Scandolo interatomic
force field approach. The
potential is optimized to reproduce the forces, energies and stresses in relaxed and strained
configurations as well as  \{0001\}, \{10\={1}0\} and \{11\={2}0\} surfaces of
Al$_{2}$O$_{3}$. Details of the force field generation are given, and its
validation is demonstrated. We apply the developed potential to investigate crack
propagation in  $\alpha$-Al$_{2}$O$_{3}$ single crystals.
\end{abstract}

\maketitle

\newcommand{\oneline}[5]{
  #1 & #2 & #3 & #4 & #5  \\}

\section{Introduction}
\label{sec:intro}

Aluminum oxide is the most commonly used ceramic in technological
applications. Due to its insulating properties, it is frequently adopted in
microelectronic devices, i.e., field effect transistors, integrated circuits or superconducting quantum interference devices.  
Another important field of application is the coating of metallic components. Alumina covering
aluminum is known to prevent further oxidation of the metal. Since
alumina is characterized by a high melting point and also a high degree of hardness,
applications at high temperatures and high mechanical demands are
possible. Mechanical properties such as tensile strengths and fracture
processes are of high importance regarding these applications. 
Because surfaces and internal interfaces play an important part in these technologies, 
numerical modelling is often focused on such systems.

Due to this technological significance, alumina and metal/alumina
interfaces were frequently investigated 
experimentally\cite{Loong1999,matsuo,evans} and by \emph{ab-initio} methods\cite{mari,zhang1,zhang2,eremeev}. 
First-principles methods are well established for calculating
the work of separation or the surface energy. However, the investigation of
dynamic processes such as crack propagation requires systems with significantly more atoms and larger timescales
than \emph{ab-initio} based methods nowadays can deal with. Hence, it is an important challenge to develop a
suitable interaction force field for classical Molecular Dynamics (MD) simulations.      

There exist interaction potential approaches for alumina, which reach a
good accuracy by using angular dependent terms \cite{vashi,blonski}. 
Pair potentials are also widely applied in simulations of alumina\cite{bodur,hoang,sun2,wunderlich} 
by reason of usually much lower computational demands. 
Due to the lattice misfit,
especially the simulation of interface structures has to be performed with many atoms. 
Hence, for the simulation of crack propagation, a pair potential is to be favored concerning
the computational effort. 

Many-body effects can be important for correctly describing bond angles and
bond-bending vibration frequencies in oxides.
However, for SiO$_{2}$\cite{itapdb:Tangney2002, Kermode2010, Beck2011} and MgO\cite{Beck2011}, it was shown
that an accurate description can be achieved without angular dependent terms, but
with polarizable oxygen atoms. There, the potential model of Tangney
and Scandolo\cite{itapdb:Tangney2002} (TS) is
applied, in which the dipole moments of oxygen atoms are determined
self-consistently from the local electric field of surrounding charges and dipoles. 

In Ref.~\onlinecite{Beck2011}, the computational effort in simulations scales 
linearly in the number of particles, which is due to the Wolf summation\cite{itapdb:Wolf1999}.
We apply this combination of the TS model and the Wolf summation method in the potential generation as well as
in simulations. An effective atomic interaction potential is generated with the program \emph{potfit}\cite{potfit}.
The potential parameters are optimized by matching the resulting forces,
energies and stresses with respective 
\emph{ab-initio} values with the force matching method\cite{itapdb:Ercolessi1994}. Simulations are performed 
with the MD code IMD\cite{imd}.

A detailed description of the adopted methods and the potential generation is given in
Sec.~\ref{sec:generation}. MD simulations and \emph{ab-initio}
calculations which were performed in order to validate the obtained force field
are presented in Sec.~\ref{sec:validation}. As a first appplication of the new
potential, we investigate crack propagation of alumina single
crystals. To our knowledge, these are the first MD simulations of crack propagation in
alumina with a potential that takes into account the polarizability of oxygen atoms.
Results of these simulations are shown in \ref{sec:results}.
Finally, conclusions and an outlook are presented in Sec.~\ref{sec:conclusion}.

\section{Force field generation}
\label{sec:generation}

\subsection{Tangney-Scandolo force field}

The TS force field is a sum of two contributions: a short-range pair
potential of Morse-Stretch (MS) form, and a long-range part, which
describes the electrostatic interactions between charges and induced
dipoles on the oxygen atoms. The MS interaction between an atom of
type $i$ and an atom of type $j$ has the form
\begin{equation}
\label{eq:MS}
  U^{\text{MS}}_{ij} =
  D_{ij}\left[\exp[\gamma_{ij}(1-\frac{r_{ij}}{r^0_{ij}})] -
    2\exp[\frac{\gamma_{ij}}{2}(1-\frac{r_{ij}}{r^0_{ij}})] \right], 
\end{equation}
with $r_{ij}=\left|\bm{r}_{ij}\right|$,
$\bm{r}_{ij}=\bm{r}_{j}-\bm{r}_{i}$ and the model parameters $D_{ij}$,
$\gamma_{ij}$ and $r^0_{ij}$, which have to be determined during the force 
field generation.

Because the dipole moments depend on the local electric field of the
surrounding charges and dipoles, a self-consistent iterative
solution has to be found. In the TS approach, a dipole moment
$\bm{p}_i^{n}$ at position $\bm{r}_i$ in iteration step $n$ consists
of an induced part due to an electric field $\bm{E}(\bm{r}_i)$ and
a short-range part $\bm{p}^{\text{SR}}_i$ due to the short-range interactions
between charges $q_i$ and $q_j$. Following Rowley
\etal\cite{itapdb:Rowley1998}, this contribution is given by
\begin{equation}
  \label{eq:TS1}
\bm{p}^{\text{SR}}_i = \alpha_i \sum\limits_{j \neq i} \frac{q_j
  \bm{r}_{ij}}{r_{ij}^3}f_{ij}(r_{ij}) 
\end{equation}
with
\begin{equation}
  \label{eq:TS2}
f_{ij}(r_{ij}) = c_{ij} \sum\limits_{k=0}^4 \frac{(b_{ij}r_{ij})^k}{k!}e^{-b_{ij}r_{ij}}.
\end{equation}
$b_{ij}$ and $c_{ij}$ are parameters of the model. Together with the
induced part, one obtains
\begin{equation}
  \label{eq:TS3}
  \bm{p}_i^n=\alpha_i\bm{E}(\bm{r}_i;\{\bm{p}_j^{n-1}\}_{j=1,N},
  \{\bm{r}_j\}_{j=1,N}) +  \bm{p}_i^{\text{SR}}, 
\end{equation}
where $\alpha_i$ is the polarizability of atom $i$ and
$\bm{E}(\bm{r}_i)$ the electric field at position $\bm{r}_i$, which is
determined by the dipole moments $\bm{p}_j$ in the previous iteration
step.
Taking into account the interactions between charges $U^{\text{qq}}$, between dipole moments $U^{\text{pp}}$
and between a charge and a dipole $U^{\text{pq}}$, the total electrostatic contribution is given by
\begin{equation}
\label{eq:Energy1}
U^{\text{EL}} = U^{\text{qq}} + U^{\text{pq}} + U^{\text{pp}},
\end{equation}
and the total interaction is
\begin{equation}
\label{eq:Energy2}
 U^{\text{tot}} = U^{\text{MS}} + U^{\text{EL}}.
\end{equation}

\subsection{Wolf summation}
 
The electrostatic energies of a condensed system are described by 
functions with $r^{-n}$ dependence, $n\in\{1,2,3\}$. For point charges 
($r^{-1}$) it is common to apply the Ewald method, where the total Coulomb 
energy of a set of N ions,
\begin{equation}
  \label{eq:Wolf1}
  E^{\text{qq}} = \frac{1}{2} \sum\limits_{i=1}^{N}
  \sum^{N}_{\substack{j=1 \\ j \neq i}}\frac{q_{i}q_{j}}{r_{ij}}, 
\end{equation}
is decomposed into two terms $E_{\bm{r}}^{\text{qq}}$ and
$E_{\bm{k}}^{\text{qq}}$ by inserting a unity of the form
$1=\erfc\!\left(\kappa r\right)+\erf\!\left(\kappa r\right)$ with the
error function
\begin{equation}
  \label{eq:Wolf2}
\erf\!\left(\kappa r\right):=\frac{2}{\sqrt{\pi}}\int\limits_0^{\kappa r}\!dt\,e^{-t^2}.
\end{equation}
The splitting parameter $\kappa$ controls the distribution of energy
between the two terms. The short-ranged erfc term 
is summed up directly, while the smooth erf term is
Fourier transformed and evaluated in reciprocal space. This restricts
the technique to periodic systems. However, the main disadvantage is
the scaling of the computational effort with the number of particles
in the simulation box, which increases as
$O(N^{3/2})$\cite{itapdb:Fincham1994}, even for the optimal choice of
$\kappa$.

Wolf \etal\cite{itapdb:Wolf1999} designed a method with linear scaling properties $O(N)$ for Coulomb
interactions. By taking into account the physical properties of the system,
the reciprocal-space term $E_{\bm{k}}^{\text{qq}}$ is disregarded. It 
can be written as
\begin{equation}
  \label{eq:Wolf3}
 E_{\bm k}^{\text{qq}} = \frac{2\pi}{V} \sum_{\substack{\bm k\ne\bm 0\\ |\bm k|<k_{c}}}
      S(\bm k)\frac{\exp(-\frac{|\bm k|^{2}}{4\kappa^{2}})}{|\bm k|^{2}}, \\
\end{equation}
where $S(\bm k)$ is the charge structure factor and
$V$ the volume of the simulation box. The charge structure factor
is the Fourier transform of the charge-charge autocorrelation
function. In condensed matter, the charges screen each other; 
the approach yields proper results in liquids and largely also in solids. 
This means that for small magnitudes of the wave vectors
$\bm{k}$, the charge structure factor is also small. In previous
studies\cite{Beck2011,Brommer2010} it was shown that the reciprocal-space term is
indeed negligible compared to the real-space part, 
provided that the splitting parameter $\kappa$ is chosen
small enough. As however a smaller $\kappa$ results in a larger
real-space cutoff $r_c$, the latter has to be chosen three or four
times the size of a typical short-range cutoff radius in metals. 
In a recent study\cite{Beck2011} a
cutoff radius of only 8 \AA\ was sufficient to describe the
electrostatics of liquid silica and magnesia accurately. 

In addition, a continuous and smooth cutoff of the remaining screened
Coulomb potential $\tilde{E}^{\text{qq}}(r_{ij})=q_iq_j\erfc(\kappa r_{ij}) r_{ij}^{-1}$
is adopted at $r_c$ by shifting the potential so that it goes to zero
smoothly in the first two derivatives at $r=r_c$. We use the Wolf
method for Coulomb and dipolar interactions. For more information about the Wolf summation 
of dipole contributions, the estimation of errors attended by 
this method and a detailed analysis of the energy conservation 
in MD simulations see Ref. \onlinecite{Brommer2010}.

\subsection{Generation with \emph{potfit}}

The programm \emph{potfit} generates an effective atomic interaction
force field solely from \emph{ab-initio} reference structures.
The potential parameters are optimized by matching the resulting
forces, energies and stresses to according first-principles values
with the force matching method. All
reference structures used in this study were prepared 
with the first-principles plane wave code VASP\cite{Kresse1993,Kresse1996}. 
For $N_m$ particles, reference configuration $m$ provides one energy
$e^0_{m}$, six components of the stress tensor $s^0_{m,l}$ ($l =
1,2,..,6$) and $3N_m$ total force cartesian components $f^0_{m,n}$ ($n =
1,2,...,3N_m$) on $N_m$ atoms. The function
\begin{equation}
\label{eq:Impl1}
Z = w_e Z_e + w_s Z_s + Z_f
\end{equation}
is minimized, where
\begin{align}
\label{eq:Impl2}
\begin{split}
 Z_e &= 3\sum\limits_{m=1}^{M} N_m(e_{m}-e^0_{m})^2, \\ 
 Z_s &= \frac{1}{2}\sum\limits_{m=1}^{M}
 \sum\limits_{l=1}^{6}N_m(s_{m,l}-s^0_{m,l})^2, \\ 
 Z_f &= \sum\limits_{m=1}^{M}\sum\limits_{n=1}^{3N_m}
 (f_{m,n}-f^0_{m,n})^2,
\end{split}
\end{align}
and $e_{m}$, $s_{m,l}$ and $f_{m,n}$ are the corresponding values calculated
with the parametrized force field. 
$w_e$ and $w_s$ are weights to balance the different amount of
available data for each quantity. They are defined in such a way, that each weight 
can be taken as the percentual amount of the concerning force data.
In the following we assume $M$ reference structures that all consist of 
the same number of particles ($N_m = N$), but in
principle, \emph{potfit} can handle any different number of particles for each
reference structure. The root mean square (rms) errors,
\begin{equation}
\label{eq:Impl3}
\Delta F_e = \sqrt{\frac{Z_e}{3MN}}~~,~~\Delta F_s = \sqrt{\frac{2Z_s}{MN}}~~~\text{and}~~\Delta F_f = \sqrt{\frac{Z_f}{MN}},
\end{equation}
are first indicators of the quality of
the generated force field. Their magnitudes are independent of
weighting factors, number and sizes of reference structures. 
For the minimization of the force field parameters, a combination of a stochastic simulated annealing
algorithm\cite{itapdb:Corona1987} and a conjugate-gradient-like deterministic
algorithm\cite{itapdb:Powell1965} is used. 
For more information about the implementation of long-range electrostatic interactions
including the TS force field with Wolf summation see Ref. \onlinecite{Beck2011}.

\subsection{Parametrization}
\label{reference}
We first prepared a set of 67 $\alpha$-alumina crystal 
structures with 360 atoms, in total 24 120 atoms. This 
reference database is composed of three kinds of structures: (i) crystals
strained up to 20\% in [0001], [\={2}110] and [0\={1}10] directions at zero Kelvin, 
(ii) structures with free (0001), (\={2}110) and (0\={1}10) surfaces at zero Kelvin and (iii) 
equilibrated snapshots taken out of an \emph{ab-initio} MD trajectory, where an
ideal crystal is heated from zero Kelvin up to 2000 Kelvin. 
It has to be mentioned that no initial ad-hoc potential as used in Refs. \onlinecite{itapdb:Tangney2002, Beck2011} is required 
to generate a reference database. This is only necessary in the case of liquid structures. The present work, however, 
deals with crystalline structures where a database can be generated without need of classical MD trajectories. 
The 67 structures are applied as input
configurations for the first-principles plane wave code VASP.\cite{Kresse1993,Kresse1996} 
We used ultrasoft pseudopotentials\cite{itapdb:Kresse1999}
and the Perdew-Wang 91 \cite{pw91} generalized gradient approximation (GGA) to the exchange-correlation functional.
The latter is recommended in Ref. \onlinecite{Beck2011}. A plane wave cutoff of
396 eV was used. We adopt a gamma
centered k mesh of 2 $\times$ 2 $\times$ 2 k points. 
The weights in \emph{potfit} were chosen to $w_e = 0.03$ and $w_s =
0.28$. Setting $\kappa= 0.1\ \text{\AA}^{-1}$, 
a cutoff radius of 10 \AA\ was found to be sufficient.
The final set of parameters is shown in Table~\ref{tab:para}.
The rms errors are $\Delta F_e = 0.0492$ eV, $\Delta F_s = 0.0273$ eV\AA$^{-3}$ and $\Delta F_f = 0.3507$ eV\AA$^{-1}$. 
\begin{table}
   \centering
   \begin{tabular}{c c c c c}
\hline
\hline
\oneline{$q_{\text{Al}}$}{$q_{\text{O}}$}{$\alpha_\text{O}$}{$b_{\text{Al}-\text{O}}$}{$c_{\text{Al}-\text{O}}$} 
\oneline{~1.122 608~}{-0.748 406}{~0.026 576~}{18.984 286}{~-5.571 329}
\hline
\oneline{}{}{$D$}{$\gamma$}{$r^0$}
\hline
\oneline{Al--Al}{}{0.000 890}{12.737 442}{5.405 175}
\oneline{Al--O}{}{1.000 058}{8.077 778}{1.851 806}
\oneline{O--O}{}{0.005 307}{12.081 851}{3.994 815}
\hline
\hline
\end{tabular}
\caption{Force field parameters, given in IMD units set eV,
  \AA\ and amu (hence charges are multiples of the elementary charge).
} 
\label{tab:para}
\end{table}
\section{Validation}
\label{sec:validation}

Firstly, we determine basic properties of crystalline alumina such as
lattice constants, cohesive energies and vibrational properties.
Additionally and relevant for fracture studies, our validation
simulations focus on surface relaxations, surface energies and stresses of
strained configurations. Finally, we probe the new potential beyond its optimization range by
investigating two basal twins.

Lattice constants and cohesive energies are obtained by pressure relaxation: Besides energy minimization, the pressure
tensor of the sample is calculated at each step, and the size of the simulation box is changed by a small
amount in order to lower that pressure. After inserting surfaces or
interfaces into the relaxed sample, a further relaxation is performed which
reveals surface or interface energies. 
The \emph{ab-initio} calculations of these
properties are performed with VASP, as described in section \ref{reference}.
As can be seen from Table~\ref{tab:valid}, the lattice constants and
the cohesive energy of crystalline $\alpha$-alumina obtained with the new potential agree well with our 
\emph{ab-initio} calculations and previous studies. %

\begin{table*}
   \centering
\begin{tabularx}{1.0\linewidth}{p{3.2cm}XXXXXXXX}
\hline
\hline
&~~a & ~~c & E$_{\text{coh}}$& $\text{E}_{\text{\{0001\}}}$
& $\text{E}_{\text{\{11\={2}0\}}}$ &$\text{E}_{\text{\{10\={1}0\}}}$ & $\text{E}_{\text{twin}}^{\text{m}}$ &  $\text{E}_{\text{twin}}^{\text{r}}$ \\
&~(\AA) & ~(\AA) &  (eV) & (J/m$^{2}$) &(J/m$^{2}$) & (J/m$^{2}$)& (J/m$^{2}$)&(J/m$^{2}$) \\
\hline
New potential & 4.79 & 12.97  & 31.85    & 1.59    & 1.65  & 1.89 & 1.32 & 0.57\\
 \hline 
Ab-initio  &  4.78 & 13.05  & 32.31  & 1.55  & 1.83  & 1.98 & 2.12 & 0.76\\
\hline
Lit. (experiment) & 4.76 \cite{pear} & 13.00 \cite{pear} & 31.8 \cite{cohes} & & & & & \\
\hline
Lit.    (ab-initio) &  &  & 33.0 \cite{siegel}  & 1.85 \cite{sun}   & 2.44 \cite{sun} & 2.39 \cite{sun} & 1.99 \cite{mari} & 0.73 \cite{mari} \\
  &  &   &   &  2.03 \cite{mack}   & 2.23 \cite{mack} & 2.50 \cite{mack} & & \\
  &  &   &     & 1.76 \cite{mana}   &   & & &\\
  &  &   &     & 1.54 \cite{pinto}   &   & & &\\
\hline
\hline
\end{tabularx}
\caption{Lattice constants, cohesive energies and
interface energies of basal twins obtained with the new potential compared to \emph{ab-initio}
results and literature data.
} 
\label{tab:valid}
\end{table*}

\begin{figure}
  \centering
  \includegraphics{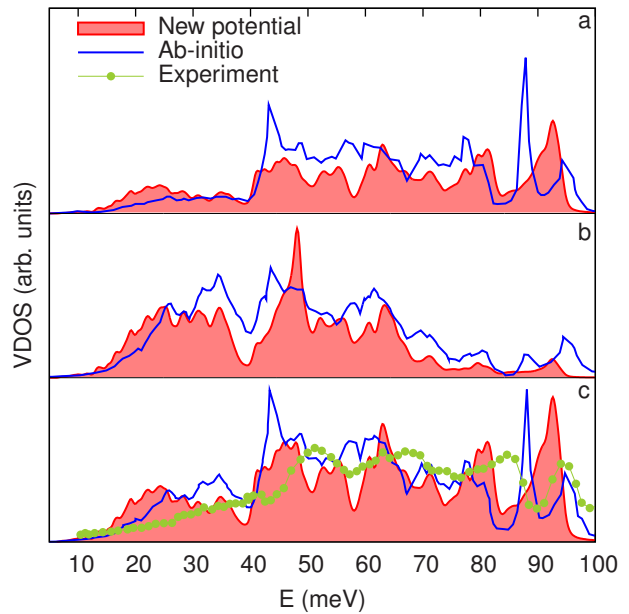}
 \caption{VDOS calculated with the new potential
   compared with an \emph{ab-initio} study\cite{Lodziana2003} and experiment\cite{Loong1999}. a) Partial VDOS for aluminum atoms,
   b) partial VDOS for oxygen atoms, c) generalized VDOS.}
  \label{fig:vdos}
\end{figure}

The \emph{partial} vibrational density of states (VDOS) for G$_{\text{Al}}$(E) and G$_{\text{O}}$(E) 
was obtained by computing the Fourier transform of the time-dependent velocity-velocity autocorrelation
function from a 100 ps MD trajectory with the software package nMoldyn\cite{Rog2003}. 
The \emph{generalized} VDOS G(E) is then calculated by
\begin{equation}
G(E) = \sum\limits_{\mu = \text{Al,O}}^{} \frac{\sigma_{\mu}}{m_{\mu}} G_{\mu}(E)
\end{equation}
with the scattering cross section $\sigma_{\mu}$ and atomic mass $m_{\mu}$ of atom $\mu$.
Fig.~\ref{fig:vdos} shows the partial and generalized VDOS. The key features of the curves obtained
with the new potential and an \emph{ab-initio} study from Ref. \onlinecite{Lodziana2003} coincide.
For the partial VDOS of aluminum, both studies show a broad band between 41 and 83 meV. The \emph{ab-initio}
results show less states in the low-energy band between 15 and 30 meV. There are two sharp peaks 
between 87 and 100 meV in the ab-inito curve, whereas the new potential shows one broader peak with a shoulder,
that indicated the second peak. The curves for the partial VDOS of oxygen are in good agreement. Both simulation
and \emph{ab-initio} calculation show the main band of states between 12 and 85 meV with similar curve progression and 
three local maxima at around 35, 47 and 62 meV. These characteristics are also reflected in the generalized
VDOS. In addition, the generalized VDOS obtained with neutron scattering\cite{Loong1999} is depicted, which 
shows the same main characteristics of the curve on a qualitative level. 

The literature surface energies\cite{sun,mack,mana,pinto} given in Table~\ref{tab:valid} differ among each
other which origins from the different methods used in
the \emph{ab-initio} approaches. Our calculation of the (0001) surface energy
agrees with the value obtained in Ref. \onlinecite{pinto},  where also the VASP
code with the same pseudopotential and exchange-correlation approximation was used. 
For all investigated surfaces, the energies obtained with the new
potential and with \emph{ab-initio} methods
agree. Both results reveal that the (0001) surface has the lowest
surface energy and the (0\={1}10) surface the highest one. 
The studies of Ref. \onlinecite{sun,mack,mana} yield slightly higher energies for all investigated surfaces.

\begin{figure}
{		
		\includegraphics[clip,angle=-0,width=0.5\linewidth]{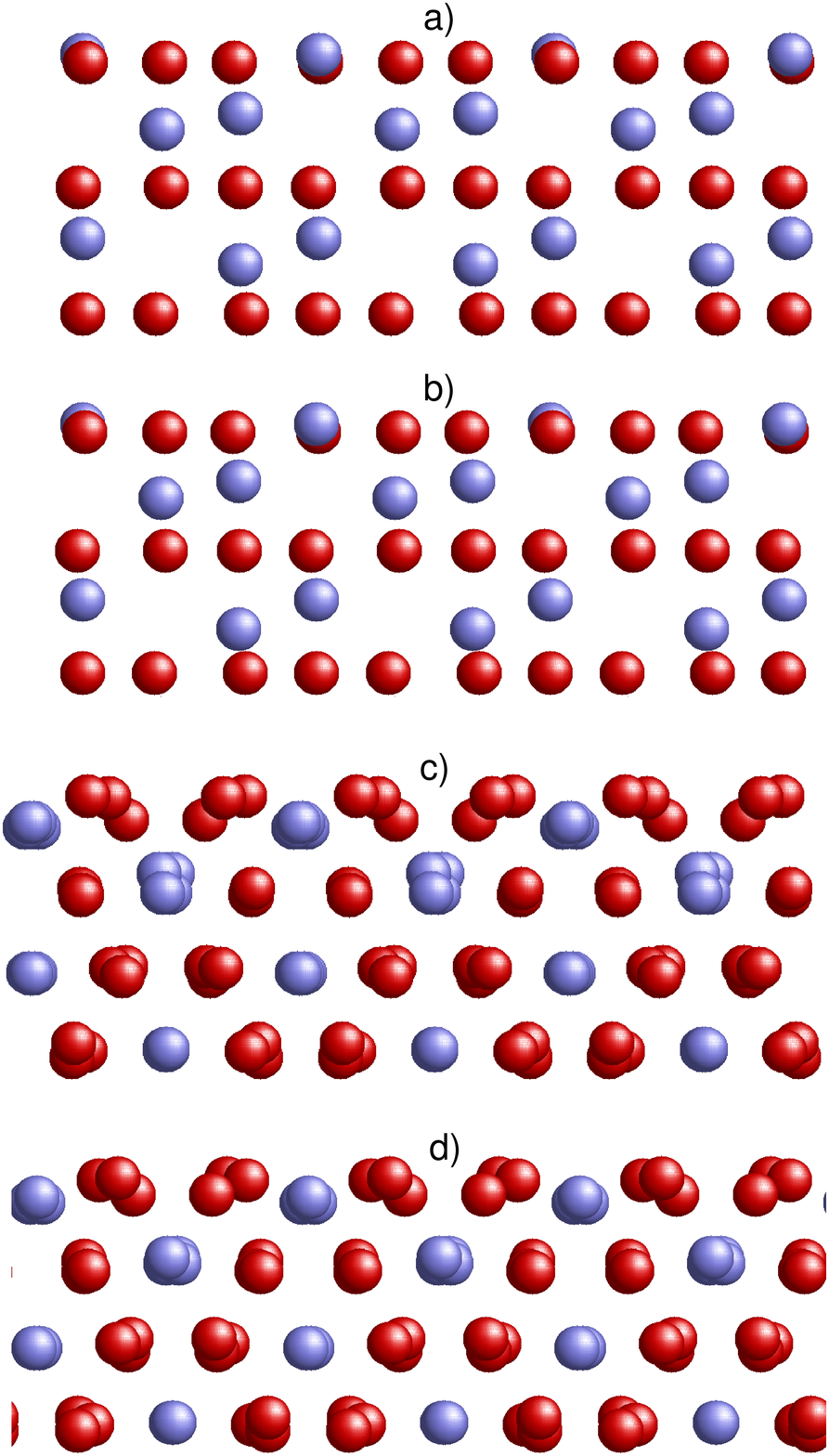}}
		\caption{Surface relaxations for the new
                  potential compared with \emph{ab-initio}
                  calculations. Red: oxygen. Blue: aluminum. a)
                  (0001)-surface, \emph{ab-inito}, b) (0001)-surface, MD, c)
                  (0\={1}10)-surface, \emph{ab-inito}, d) (0\={1}10)-surface, MD.
                }
		\label{fig:rel}
	\end{figure}

The surface structure after relaxation is shown in Fig.~\ref{fig:rel}. It reveals
that the Al atoms of the outermost Al-layer are moved closer to the outermost O-layer
at the (0001) Al-terminated surface, which is known to be the most stable
(0001) surface termination. The atomic adjustment perpendicular to the surface obtained 
with the new potential agrees very well with our \emph{ab-initio} results.
The distance of an Al- and an O-layer is 0.83 \AA, whereas this value decreases to 0.15
\AA\,(MD) respectively 0.14  \AA\,(\emph{ab-initio}) at the surface. 
A relaxation of the oxygen atoms can be seen at the (0\={1}10) surface.
Both MD and \emph{ab-initio} study show that 
the three oxygen atoms per unit cell -- which are initially in a row along the
direction orthogonal to the plane of Fig.~\ref{fig:rel} -- relax to different distances from the
initial surface. Furthermore, a relaxation towards the neighboring Al-atoms in the first layer occurs.
The results obtained with the new potential again coincide with the \emph{ab-initio} calculation.
One difference, however, can be seen at the second layer: Every second
Al atom in each row orthogonal to the figure plane is slightly displaced towards the surface in the
\emph{ab-initio} relaxation. This effect is not observed in the MD relaxation simulation. 
Neither the MD nor the \emph{ab-inito} relaxation study of the (\={2}110) surface yield significant atomic movements.

As described in section \ref{reference}, strained configurations with strain
directions [0001], [\={2}110] and [0\={1}10] were added in the
reference database for the potential optimization approach. To clarify,
whether the new potential can reproduce the stresses of strained
configurations, these stresses are calculated using MD with the new
potential. 
 
\begin{figure}
{		
		\includegraphics[angle=-0]{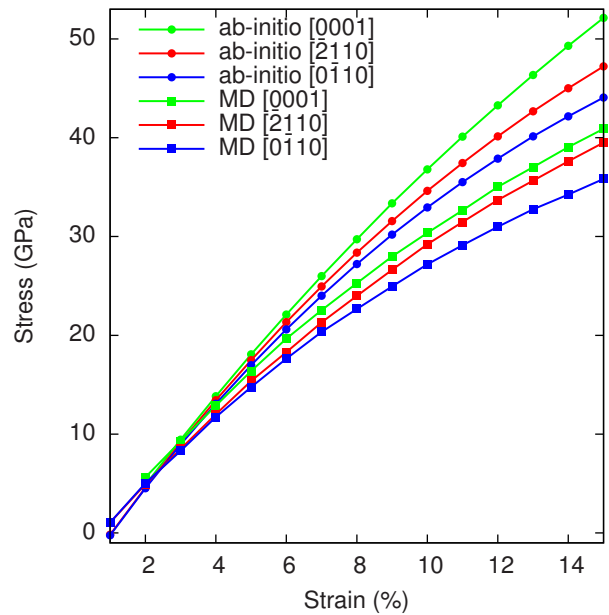}}
		\caption{Stresses of strained configurations with [0001] (green),
                  [\={2}110] (red) and [0\={1}10] (blue) strain directions against the
                  stress tensor component in direction of strain. The curves obtained with the
		  new potential are marked with squares, those from \emph{ab-initio} calculations
		  with circles.
}
		\label{fig:stress}
	\end{figure}

Fig.~\ref{fig:stress} shows a comparison of stresses obtained in simulations
to the stresses of the
underlying \emph{ab-initio} reference configurations, each with 360 atoms.
They are strained up to 15\% in [0001], [\={2}110] and [0\={1}10] direction respectively. 
The difference between MD and \emph{ab-initio} results is small at
lower strains. With increasing strain, the difference increases up to
about 10 GPa at 15\% strain. The new potential underestimates the stress in all cases.
However, the directions, in which the stress increases, can be reproduced correctly:
The highest stresses are observed for strains in [0001] direction, the
lowest stresses are found for configurations strained in [0\={1}10]
direction.

Finally, we probe the potential by simulating basal twins. These are systems which were not
included in the optimization process. This demonstrates the applicability of the new force field
beyond the range for which it was optimized.
It is well known that
the basal twin is a frequently observed Al$_{2}$O$_{3}$ interface\cite{morr}. 
We consider two types of basal twins which differ by their
interface structure: The mirror twin consists of two single crystal
elementary cells with mirror symmetry. The mirror plane is a (0001) oriented
oxygen layer. Based on the mirror twin, one can obtain the rotation twin
by shifting one of the single crystal elementary cells in the [0\={1}10]
direction. For details of the interface structures see Ref. \onlinecite{mari}. As
can be seen from Table~\ref{tab:valid}, the \emph{ab-inito} calculated interface energies of the basal
twins agree well with the results from Ref. \onlinecite{mari}. The MD simulations with the new potential
underestimate both energy values. But the potential is able to qualitatively reproduce the relation of
the two interface energies: The interface energy of the mirror twin ($\text{E}_{\text{twin}}^{\text{m}}$) is about two by three times the one
of the rotation twin ($\text{E}_{\text{twin}}^{\text{r}}$); that is why, the latter one is energetically favorable. 
\section{Crack propagation}
\label{sec:results}
\subsection{Simulation conditions}

To investigate crack propagation in mode I at constant energy release rate,
we prepared configurations with dimensions ($b_x, b_y, b_z$) of 
$21 \times 3 \times 13$ nm$^3$ which contain about 80 000 atoms. Lateral planes of these cuboids are the (0001),
(0\={1}10) and (\={2}110) crystallographic planes.
An elliptical initial crack of 5 nm length in x-direction is inserted on one side of the
samples by moving atoms in z-direction. The opening of the crack is calculated with the Griffith criterion: In front of the crack tip the
sample is linearly strained until the elastic energy due to strain is equal
to the Griffith energy G$_{0}$ which is twice the surface energy of the crack plane. 
This criterion was fullfilled at
about 4\% strain. These structures are then relaxed to
obtain the displacement field of a stable crack. Periodic boundary conditions are applied in the
direction along the crack front, whereas fixed displacement boundary conditions are
applied in the other directions. Initial configurations for
crack propagation simulations with different energy release rates are obtained by a linear
scaling of this displacement field. During the crack propagation simulations, the NVE 
(constant number of particles, volume and energy) 
ensemble and a starting temperature of 0 K are applied. A timestep of 1$\,$fs
is used for all simulations.

	\begin{figure}
{		
		\includegraphics[angle=-0,width=0.5\linewidth]{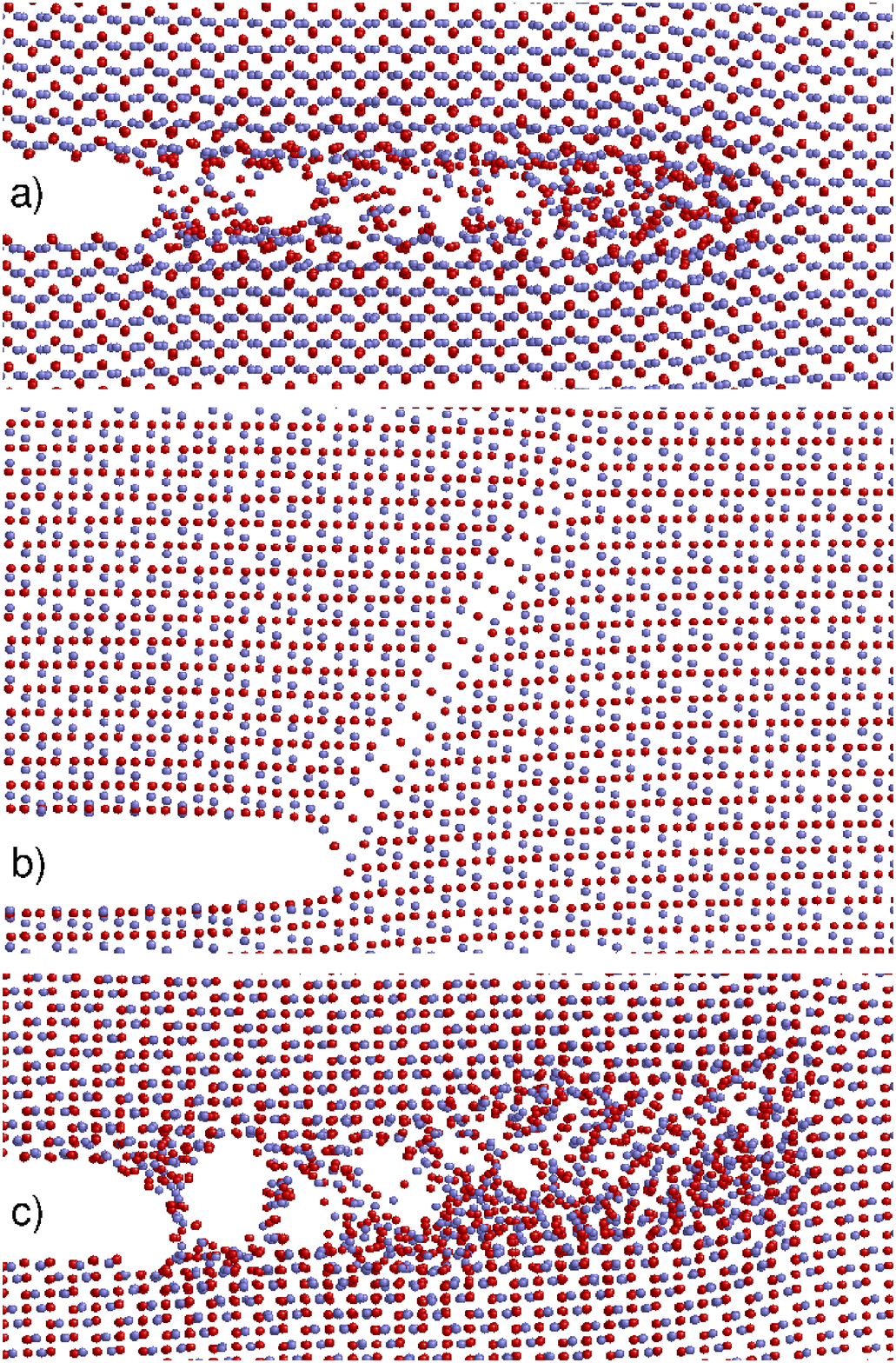}}
		\caption{Crack propagation in $\alpha$-Al$_{2}$O$_{3}$ (Al-atoms
                  blue, O-atoms red). a)
                 Initial crack in (\={2}110) plane, b) initial crack in
                  (0001) plane, c) initial crack in
                  (0\={1}10) plane.                 
                }
		\label{fig:riss1}
	\end{figure}

\subsection{Cracks in a \{11\={2}0\} plane}

Initial cracks were inserted in a (\={2}110) plane with crack propagation
directions [0001] or [0\={1}10]. In both cases, the initial crack is stationary at
energy release rates up to 1.5$\,$G$_{0}$. Higher energy release rates lead to crack
propagation in the initial (\={2}110) plane. It is known that the discrete character of
interatomic bonds which reveals itself in the so called lattice trapping
effect \cite{lattice} can retard crack propagation to energy release rates beyond
the critical energy release rate determined by the Griffith criterion. 
As can be seen from Fig.~\ref{fig:riss1} a, yet at high energy release rates bond breaking between the two crack
lips is not continuous. Chains of atoms bridging the crack lips stay intact
during the simulation. Furthermore, considerable disorder is observed at the
resulting crack surfaces. In [0\={1}10] propagation direction, these effects
are slightly more pronounced than in [0001] propagation direction. It
cannot be ruled out that these very strong bonds are caused by inaccuracies of
the force field. However, the observation of intact bonds bridging the crack lips  
agree with previous simulations of alumina, where it was stated that these atomic chains were observed in
experiments as well \cite{wunderlich}. 
Our results show that cracks are able to propagate in (\={2}110) planes in both
[0001] and [0\={1}10] directions. This is in accordance with experiments \cite{matsuo}, where it turned out
that cracks occur in \{11\={2}0\} planes although these are not the preferred fracture
planes.

\subsection{Cracks in a \{0001\} plane}

The \{0001\} planes are the closest packed planes of $\alpha$-Al$_{2}$O$_{3}$;
therefore cleavage of these planes is improbable. Our simulations confirm that cracks do not propagate in
the (0001) plane in which the initial crack was inserted with crack propagation
directions [\={2}110] or [0\={1}10]. At energy release
rates below 1.7$\,$G$_{0}$, a damaged
region in front of the crack tip is generated with atomic disorder similar to
the case of the (\={2}110) plane. However, the crack tip does not move. More
interesting is the (0001)[0\={1}10] oriented initial crack at higher energy release rates: As can be seen from
Fig.~\ref{fig:riss1} b, the crack propagates in a \{10\={1}2\} cleavage
plane. Due to the boundary conditions, the crack surfaces cannot separate
completely, but a row of oxygen atoms stays in the crack path. In contrast to
the cracks in a \{11\={2}0\} plane, there is no disorder along the crack
surfaces. The observed crack propagation in a \{10\={1}2\} cleavage
plane agrees well with electron microscopy investigations
\cite{matsuo} which revealed that fracture surfaces of the \{10\={1}2\}
cleavage planes are frequently observed in alumina.  

\subsection{Cracks in a \{10\={1}0\} plane}

Cracks initially inserted in a (0\={1}10) plane in [\={2}110] direction
propagate in this orientation starting at energy release
rates of 2.2$\,$G$_{0}$. In the case of a [0001] crack propagation direction
(Fig.~\ref{fig:riss1} c), the movement of the crack tip was
observed at energy release rates above 1.9$\,$G$_{0}$. Similar to the case of
the (0001)[0\={1}10] orientation, the crack changes its propagation plane. As can
be seen in Fig.~\ref{fig:riss1} c, the crack moves partially in
the initial (0\={1}10) plane, but also partially in a \{10\={1}2\} cleavage
plane. As in the case of cracks in \{11\={2}0\} planes, the initially
(0\={1}10)[\={2}110] or (0\={1}10)[0001] oriented cracks generate disorder at the crack
surfaces during crack propagation; some atomic bonds stay intact across the
crack opening.

\subsection{Orientation of electric dipole moments}

	\begin{figure*}
{		
		\includegraphics[angle=-0,width=1\linewidth]{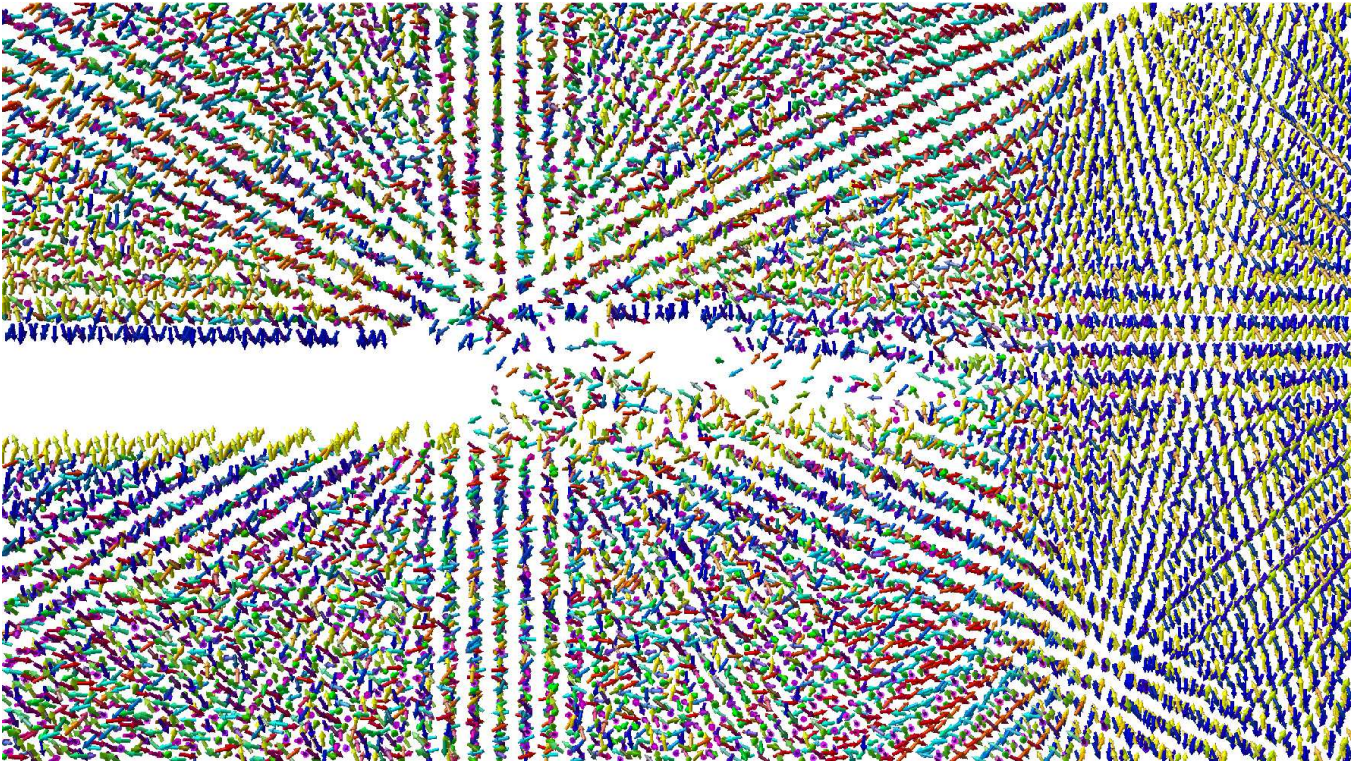}}
		\caption{Electric dipole moments of oxygen atoms in crack propagation
                  simulation. Each oxygen atom is visualized by an arrow with the
                  direction of the dipole moment; the lengths are normalized. The colour corresponds to the
                  orientation: blue -- down,  yellow -- up, green -- left, red -- right.                
                }
		\label{fig:dipoles}
	\end{figure*}

The orientation of electric dipole moments in crack propagation simulations can differ significantly
from the orientations in other configurations of the
same material. There are two main effects which influence the alignment of dipoles: Charged surfaces and
piezoelectricity. In order to investigate these effects in
$\alpha$-Al$_{2}$O$_{3}$, dipoles of the crack propagation simulations are
visualized with the program MegaMol\cite{megamol}.
Since effects from the boundary conditions
(which involve unnatural surfaces in some directions) should be excluded, we performed a crack propagation
simulation with periodic boundary conditions in all directions. In this case, a
symmetrical crack was inserted in the middle of the sample. As long as the two
crack tips are far away from each other (which is true up to about 5 nm
distance), we observe the same result as with the boundary conditions described
above. Hence, possible boundary condition effects can be excluded.

Charged crack surfaces occur as a function of
the crack propagation planes. As shown in Fig.~\ref{fig:riss1}, the
initial surfaces of cracks in a \{11\={2}0\} plane are oxygen terminated. Therefore, it can be expected that
the dipoles of the oxygen atoms at each crack surface are aligned in the same direction and the
alignment on the other crack surface is the other way round.
This adjustment is reproduced in our simulation, as can be seen in Fig.~\ref{fig:dipoles}.
It is significantly more pronounced at the surfaces of the
initial crack (left part of the crack in Fig.~\ref{fig:dipoles}), but it can be seen also along the crack surfaces created by the
propagation (right part of the crack). In the latter case,
the disorder along the crack surfaces and the intact atomic chains across the crack opening reduce the alignment.       
 
Collective orientation mechanisms of electric dipole moments due to strain can only be observed in crystals,
where the inversion symmetry is broken. Therefore, in bulk $\alpha$-Al$_{2}$O$_{3}$ usually no
dipole alignment is observed. With the new force field, we simulated 
$\alpha$-Al$_{2}$O$_{3}$ strained in [\={2}110] direction up to 15\%.
No collective alignment of dipole moments occured in this simulation. 
In the crack propagation simulation, however,
the inversion symmetry is broken by the crack. As can bee seen from
Fig.~\ref{fig:dipoles}, an orientation of the electric dipole moments due to
strain is observed. Dipoles are
aligned in the region in front of the crack tip which corresponds to the
strained part of the configuration. On the contrary, in the unstrained regions below and
above the crack surfaces the dipoles show no preferred orientation.      
\section{Conclusions}
\label{sec:conclusion}

We presented the generation and validation of an
effective \emph{ab-initio} based polarizable force field for MD simulations of $\alpha$-Al$_{2}$O$_{3}$.
It was shown that it can be used to simulate mechanical and vibrational properties of crystalline alumina
with high accuracy. Furthermore, the potential is designed for the simulation of surfaces, notably for
crack propagation. Due to the Wolf summation, we achieve linear scaling properties in the number of particles, 
which makes it possible to investigate typical required system sizes and time scales
for crack simulations in resonable real time.
The polarizability of oxygen atoms is taken into account by use of the Tangney-Scandolo interatomic force field approach.
To our \linebreak knowledge, the present work is the first study of the
behavior of electric dipole moments in MD simulations on crack propagation. 

As a first application beyond the study of basic crystalline properties, 
we performed crack propagation simulations with initial cracks in different
crystallographic planes of $\alpha$-Al$_{2}$O$_{3}$.
We have shown that cracks usually propagate in the initial plane, but 
no crack propagation occurs in the close packed \{0001\} planes. It was
also observed that cracks tend to deflect to a \{10\={1}2\} cleavage plane in the case
of an initial crack plane which is unfavourable regarding crack propagation.  
The simulation result, that \{10\={1}2\} cleavage planes are favourable regarding crack
propagation, agrees well with electron microscopy investigations\cite{matsuo} of cracks in
$\alpha$-Al$_{2}$O$_{3}$. Additionally, it was shown that the new force field
allows to investigate electric dipole orientations in various strained
structures. Dipole alignment due to strain in $\alpha$-Al$_{2}$O$_{3}$ was
found in the performed crack propagation simulations.    

In the future, the new force field can be applied
for large scale simulations of alumina systems, particularly with regard to
surfaces and interfaces, e.g. polycrystals, or multi-layer composites. 
The new force field allows further increasing of system sizes to several millions of atoms at time scales up
to 1\,ns. 

\begin{acknowledgments}
  Support from the DFG through Collaborative Research Centre 716, Project B.1 and B.2 is gratefully acknowledged.
The authors thank Holger Euchner for helpful discussions concerning vibrational properties of condensed materials,
and Sebastian Grottel for the collaboration regarding the visualization with MegaMol.
\end{acknowledgments}


\begin{thebibliography}{41}%
\makeatletter
\providecommand \@ifxundefined [1]{%
 \@ifx{#1\undefined}
}%
\providecommand \@ifnum [1]{%
 \ifnum #1\expandafter \@firstoftwo
 \else \expandafter \@secondoftwo
 \fi
}%
\providecommand \@ifx [1]{%
 \ifx #1\expandafter \@firstoftwo
 \else \expandafter \@secondoftwo
 \fi
}%
\providecommand \natexlab [1]{#1}%
\providecommand \enquote  [1]{``#1''}%
\providecommand \bibnamefont  [1]{#1}%
\providecommand \bibfnamefont [1]{#1}%
\providecommand \citenamefont [1]{#1}%
\providecommand \href@noop [0]{\@secondoftwo}%
\providecommand \href [0]{\begingroup \@sanitize@url \@href}%
\providecommand \@href[1]{\@@startlink{#1}\@@href}%
\providecommand \@@href[1]{\endgroup#1\@@endlink}%
\providecommand \@sanitize@url [0]{\catcode `\\12\catcode `\$12\catcode
  `\&12\catcode `\#12\catcode `\^12\catcode `\_12\catcode `\%12\relax}%
\providecommand \@@startlink[1]{}%
\providecommand \@@endlink[0]{}%
\providecommand \url  [0]{\begingroup\@sanitize@url \@url }%
\providecommand \@url [1]{\endgroup\@href {#1}{\urlprefix }}%
\providecommand \urlprefix  [0]{URL }%
\providecommand \Eprint [0]{\href }%
\providecommand \doibase [0]{http://dx.doi.org/}%
\providecommand \selectlanguage [0]{\@gobble}%
\providecommand \bibinfo  [0]{\@secondoftwo}%
\providecommand \bibfield  [0]{\@secondoftwo}%
\providecommand \translation [1]{[#1]}%
\providecommand \BibitemOpen [0]{}%
\providecommand \bibitemStop [0]{}%
\providecommand \bibitemNoStop [0]{.\EOS\space}%
\providecommand \EOS [0]{\spacefactor3000\relax}%
\providecommand \BibitemShut  [1]{\csname bibitem#1\endcsname}%
\let\auto@bib@innerbib\@empty
%
\bibitem [{\citenamefont {{Loong}}(1999)}]{Loong1999}%
  \BibitemOpen
  \bibfield  {author} {\bibinfo {author} {\bibfnamefont {C.-K.}\ \bibnamefont
  {{Loong}}},\ }\href@noop {} {\bibfield  {journal} {\bibinfo  {journal}
  {J.~Eur.\ Cer.\ Soc.}\ }\textbf {\bibinfo {volume} {19}},\ \bibinfo {pages}
  {2241} (\bibinfo {year} {1999})}\BibitemShut {NoStop}%
\bibitem [{\citenamefont {{Matsuo}}\ \emph {et~al.}(2010)\citenamefont
  {{Matsuo}}, \citenamefont {{Mitsuhara}}, \citenamefont {{Ikeda}},
  \citenamefont {{Hata}},\ and\ \citenamefont {{Nakashima}}}]{matsuo}%
  \BibitemOpen
  \bibfield  {author} {\bibinfo {author} {\bibfnamefont {H.}~\bibnamefont
  {{Matsuo}}}, \bibinfo {author} {\bibfnamefont {M.}~\bibnamefont
  {{Mitsuhara}}}, \bibinfo {author} {\bibfnamefont {K.}~\bibnamefont
  {{Ikeda}}}, \bibinfo {author} {\bibfnamefont {S.}~\bibnamefont {{Hata}}}, \
  and\ \bibinfo {author} {\bibfnamefont {H.}~\bibnamefont {{Nakashima}}},\
  }\href@noop {} {\bibfield  {journal} {\bibinfo  {journal} {Int. J. Fatigue}\
  }\textbf {\bibinfo {volume} {32}},\ \bibinfo {pages} {592} (\bibinfo {year}
  {2010})}\BibitemShut {NoStop}%
\bibitem [{\citenamefont {G.{Evans}}\ \emph {et~al.}(1986)\citenamefont
  {G.{Evans}}, \citenamefont {{Lu}}, \citenamefont {{Schmauder}},\ and\
  \citenamefont {{R{\"u}hle}}}]{evans}%
  \BibitemOpen
  \bibfield  {author} {\bibinfo {author} {\bibfnamefont {A.}~\bibnamefont
  {G.{Evans}}}, \bibinfo {author} {\bibfnamefont {M.~C.}\ \bibnamefont {{Lu}}},
  \bibinfo {author} {\bibfnamefont {S.}~\bibnamefont {{Schmauder}}}, \ and\
  \bibinfo {author} {\bibfnamefont {M.}~\bibnamefont {{R{\"u}hle}}},\
  }\href@noop {} {\bibfield  {journal} {\bibinfo  {journal} {Acta Metall.}\
  }\textbf {\bibinfo {volume} {34}},\ \bibinfo {pages} {1643} (\bibinfo {year}
  {1986})}\BibitemShut {NoStop}%
\bibitem [{\citenamefont {{Marinopoulos}}, \citenamefont {{Nufer}},\ and\
  \citenamefont {{Elsässer}}(2001)}]{mari}%
  \BibitemOpen
  \bibfield  {author} {\bibinfo {author} {\bibfnamefont {A.~G.}\ \bibnamefont
  {{Marinopoulos}}}, \bibinfo {author} {\bibfnamefont {S.}~\bibnamefont
  {{Nufer}}}, \ and\ \bibinfo {author} {\bibfnamefont {C.}~\bibnamefont
  {{Elsässer}}},\ }\href@noop {} {\bibfield  {journal} {\bibinfo  {journal}
  {Phys.\ Rev.~B}\ }\textbf {\bibinfo {volume} {63}},\ \bibinfo {pages}
  {165112} (\bibinfo {year} {2001})}\BibitemShut {NoStop}%
\bibitem [{\citenamefont {{Zhang}}\ and\ \citenamefont
  {{Smith}}(2000{\natexlab{a}})}]{zhang1}%
  \BibitemOpen
  \bibfield  {author} {\bibinfo {author} {\bibfnamefont {W.}~\bibnamefont
  {{Zhang}}}\ and\ \bibinfo {author} {\bibfnamefont {J.~R.}\ \bibnamefont
  {{Smith}}},\ }\href@noop {} {\bibfield  {journal} {\bibinfo  {journal}
  {Phys.\ Rev.\ Lett.}\ }\textbf {\bibinfo {volume} {85}},\ \bibinfo {pages}
  {3225} (\bibinfo {year} {2000}{\natexlab{a}})}\BibitemShut {NoStop}%
\bibitem [{\citenamefont {{Zhang}}\ and\ \citenamefont
  {{Smith}}(2000{\natexlab{b}})}]{zhang2}%
  \BibitemOpen
  \bibfield  {author} {\bibinfo {author} {\bibfnamefont {W.}~\bibnamefont
  {{Zhang}}}\ and\ \bibinfo {author} {\bibfnamefont {J.~R.}\ \bibnamefont
  {{Smith}}},\ }\href@noop {} {\bibfield  {journal} {\bibinfo  {journal}
  {Phys.\ Rev.~B}\ }\textbf {\bibinfo {volume} {61}},\ \bibinfo {pages} {16883}
  (\bibinfo {year} {2000}{\natexlab{b}})}\BibitemShut {NoStop}%
\bibitem [{\citenamefont {{Eremeev}}\ \emph {et~al.}(2009)\citenamefont
  {{Eremeev}}, \citenamefont {{Schmauder}}, \citenamefont {{Hocker}},\ and\
  \citenamefont {{Kulkova}}}]{eremeev}%
  \BibitemOpen
  \bibfield  {author} {\bibinfo {author} {\bibfnamefont {S.~V.}\ \bibnamefont
  {{Eremeev}}}, \bibinfo {author} {\bibfnamefont {S.}~\bibnamefont
  {{Schmauder}}}, \bibinfo {author} {\bibfnamefont {S.}~\bibnamefont
  {{Hocker}}}, \ and\ \bibinfo {author} {\bibfnamefont {S.~E.}\ \bibnamefont
  {{Kulkova}}},\ }\href@noop {} {\bibfield  {journal} {\bibinfo  {journal}
  {Physica B}\ }\textbf {\bibinfo {volume} {404}},\ \bibinfo {pages} {2065}
  (\bibinfo {year} {2009})}\BibitemShut {NoStop}%
\bibitem [{\citenamefont {{Vashishta}}\ \emph {et~al.}(2008)\citenamefont
  {{Vashishta}}, \citenamefont {{Kalia}}, \citenamefont {{Nakano}},\ and\
  \citenamefont {{Rino}}}]{vashi}%
  \BibitemOpen
  \bibfield  {author} {\bibinfo {author} {\bibfnamefont {P.}~\bibnamefont
  {{Vashishta}}}, \bibinfo {author} {\bibfnamefont {R.~K.}\ \bibnamefont
  {{Kalia}}}, \bibinfo {author} {\bibfnamefont {A.}~\bibnamefont {{Nakano}}}, \
  and\ \bibinfo {author} {\bibfnamefont {J.~P.}\ \bibnamefont {{Rino}}},\
  }\href@noop {} {\bibfield  {journal} {\bibinfo  {journal} {J. Appl. Phys.}\
  }\textbf {\bibinfo {volume} {103}},\ \bibinfo {pages} {083504} (\bibinfo
  {year} {2008})}\BibitemShut {NoStop}%
\bibitem [{\citenamefont {{Blonski}}\ and\ \citenamefont
  {{Garofalini}}(1993)}]{blonski}%
  \BibitemOpen
  \bibfield  {author} {\bibinfo {author} {\bibfnamefont {S.}~\bibnamefont
  {{Blonski}}}\ and\ \bibinfo {author} {\bibfnamefont {S.~H.}\ \bibnamefont
  {{Garofalini}}},\ }\href@noop {} {\bibfield  {journal} {\bibinfo  {journal}
  {Surf. Sci.}\ }\textbf {\bibinfo {volume} {295}},\ \bibinfo {pages} {263}
  (\bibinfo {year} {1993})}\BibitemShut {NoStop}%
\bibitem [{\citenamefont {{Bodur}}, \citenamefont {{Chang}},\ and\
  \citenamefont {{Argon}}(2005)}]{bodur}%
  \BibitemOpen
  \bibfield  {author} {\bibinfo {author} {\bibfnamefont {C.~T.}\ \bibnamefont
  {{Bodur}}}, \bibinfo {author} {\bibfnamefont {J.}~\bibnamefont {{Chang}}}, \
  and\ \bibinfo {author} {\bibfnamefont {A.~S.}\ \bibnamefont {{Argon}}},\
  }\href@noop {} {\bibfield  {journal} {\bibinfo  {journal} {J. Eur. Cer.
  Soc.}\ }\textbf {\bibinfo {volume} {25}},\ \bibinfo {pages} {1431} (\bibinfo
  {year} {2005})}\BibitemShut {NoStop}%
\bibitem [{\citenamefont {{Van Hoang}}(2004)}]{hoang}%
  \BibitemOpen
  \bibfield  {author} {\bibinfo {author} {\bibfnamefont {V.}~\bibnamefont {{Van
  Hoang}}},\ }\href@noop {} {\bibfield  {journal} {\bibinfo  {journal} {Phys.\
  Rev.~B}\ }\textbf {\bibinfo {volume} {70}},\ \bibinfo {pages} {134204}
  (\bibinfo {year} {2004})}\BibitemShut {NoStop}%
\bibitem [{\citenamefont {{Sun}}, \citenamefont {{Stirner}},\ and\
  \citenamefont {{Matthews}}(2007)}]{sun2}%
  \BibitemOpen
  \bibfield  {author} {\bibinfo {author} {\bibfnamefont {J.}~\bibnamefont
  {{Sun}}}, \bibinfo {author} {\bibfnamefont {T.}~\bibnamefont {{Stirner}}}, \
  and\ \bibinfo {author} {\bibfnamefont {A.}~\bibnamefont {{Matthews}}},\
  }\href@noop {} {\bibfield  {journal} {\bibinfo  {journal} {Surf. Sci.}\
  }\textbf {\bibinfo {volume} {601}},\ \bibinfo {pages} {1358} (\bibinfo {year}
  {2007})}\BibitemShut {NoStop}%
\bibitem [{\citenamefont {{Wunderlich}}\ and\ \citenamefont
  {{Awaji}}(2001)}]{wunderlich}%
  \BibitemOpen
  \bibfield  {author} {\bibinfo {author} {\bibfnamefont {W.}~\bibnamefont
  {{Wunderlich}}}\ and\ \bibinfo {author} {\bibfnamefont {H.}~\bibnamefont
  {{Awaji}}},\ }\href@noop {} {\bibfield  {journal} {\bibinfo  {journal}
  {Material \& Design}\ }\textbf {\bibinfo {volume} {22}},\ \bibinfo {pages}
  {53} (\bibinfo {year} {2001})}\BibitemShut {NoStop}%
\bibitem [{\citenamefont {{Tangney}}\ and\ \citenamefont
  {{Scandolo}}(2002)}]{itapdb:Tangney2002}%
  \BibitemOpen
  \bibfield  {author} {\bibinfo {author} {\bibfnamefont {P.}~\bibnamefont
  {{Tangney}}}\ and\ \bibinfo {author} {\bibfnamefont {S.}~\bibnamefont
  {{Scandolo}}},\ }\href@noop {} {\bibfield  {journal} {\bibinfo  {journal}
  {J.~Chem.\ Phys.}\ }\textbf {\bibinfo {volume} {117}},\ \bibinfo {pages}
  {8898} (\bibinfo {year} {2002})}\BibitemShut {NoStop}%
\bibitem [{\citenamefont {{Kermode}}\ \emph {et~al.}(2010)\citenamefont
  {{Kermode}}, \citenamefont {{Cereda}}, \citenamefont {{Tangney}},\ and\
  \citenamefont {{De Vita}}}]{Kermode2010}%
  \BibitemOpen
  \bibfield  {author} {\bibinfo {author} {\bibfnamefont {J.~R.}\ \bibnamefont
  {{Kermode}}}, \bibinfo {author} {\bibfnamefont {S.}~\bibnamefont {{Cereda}}},
  \bibinfo {author} {\bibfnamefont {P.}~\bibnamefont {{Tangney}}}, \ and\
  \bibinfo {author} {\bibfnamefont {A.}~\bibnamefont {{De Vita}}},\ }\href@noop
  {} {\bibfield  {journal} {\bibinfo  {journal} {J.~Chem.\ Phys.}\ }\textbf
  {\bibinfo {volume} {133}},\ \bibinfo {pages} {094102} (\bibinfo {year}
  {2010})}\BibitemShut {NoStop}%
\bibitem [{\citenamefont {{Beck}}\ \emph {et~al.}(2011)\citenamefont {{Beck}},
  \citenamefont {{Brommer}}, \citenamefont {{Roth}},\ and\ \citenamefont
  {{Trebin}}}]{Beck2011}%
  \BibitemOpen
  \bibfield  {author} {\bibinfo {author} {\bibfnamefont {P.}~\bibnamefont
  {{Beck}}}, \bibinfo {author} {\bibfnamefont {P.}~\bibnamefont {{Brommer}}},
  \bibinfo {author} {\bibfnamefont {J.}~\bibnamefont {{Roth}}}, \ and\ \bibinfo
  {author} {\bibfnamefont {H.-R.}\ \bibnamefont {{Trebin}}},\ }\href@noop {}
  {\bibfield  {journal} {\bibinfo  {journal} {J.~Chem.\ Phys.}\ }\textbf
  {\bibinfo {volume} {135}},\ \bibinfo {pages} {234512} (\bibinfo {year}
  {2011})}\BibitemShut {NoStop}%
\bibitem [{\citenamefont {{Wolf}}\ \emph {et~al.}(1999)\citenamefont {{Wolf}},
  \citenamefont {{Keblinski}}, \citenamefont {{Phillpot}},\ and\ \citenamefont
  {{Eggebrecht}}}]{itapdb:Wolf1999}%
  \BibitemOpen
  \bibfield  {author} {\bibinfo {author} {\bibfnamefont {D.}~\bibnamefont
  {{Wolf}}}, \bibinfo {author} {\bibfnamefont {P.}~\bibnamefont {{Keblinski}}},
  \bibinfo {author} {\bibfnamefont {S.~R.}\ \bibnamefont {{Phillpot}}}, \ and\
  \bibinfo {author} {\bibfnamefont {J.}~\bibnamefont {{Eggebrecht}}},\
  }\href@noop {} {\bibfield  {journal} {\bibinfo  {journal} {J.~Chem.\ Phys.}\
  }\textbf {\bibinfo {volume} {110}},\ \bibinfo {pages} {8254} (\bibinfo {year}
  {1999})}\BibitemShut {NoStop}%
\bibitem [{\citenamefont {{Brommer}}\ and\ \citenamefont
  {{Gähler}}(2007)}]{potfit}%
  \BibitemOpen
  \bibfield  {author} {\bibinfo {author} {\bibfnamefont {P.}~\bibnamefont
  {{Brommer}}}\ and\ \bibinfo {author} {\bibfnamefont {F.}~\bibnamefont
  {{Gähler}}},\ }\href@noop {} {\bibfield  {journal} {\bibinfo  {journal}
  {Modelling Simul. Mater. Sci. Eng.}\ }\textbf {\bibinfo {volume} {15}},\
  \bibinfo {pages} {295} (\bibinfo {year} {2007})},\ \bibinfo {note} {{\tt
  http://www.itap.physik.uni-stuttgart.de/\~{}imd/potfit/}}\BibitemShut
  {NoStop}%
\bibitem [{\citenamefont {{Ercolessi}}\ and\ \citenamefont
  {{Adams}}(1994)}]{itapdb:Ercolessi1994}%
  \BibitemOpen
  \bibfield  {author} {\bibinfo {author} {\bibfnamefont {F.}~\bibnamefont
  {{Ercolessi}}}\ and\ \bibinfo {author} {\bibfnamefont {J.~B.}\ \bibnamefont
  {{Adams}}},\ }\href@noop {} {\bibfield  {journal} {\bibinfo  {journal}
  {Europhys.\ Lett.}\ }\textbf {\bibinfo {volume} {26}},\ \bibinfo {pages}
  {583} (\bibinfo {year} {1994})}\BibitemShut {NoStop}%
\bibitem [{\citenamefont {{Stadler}}, \citenamefont {{Mikulla}},\ and\
  \citenamefont {{Trebin}}(1997)}]{imd}%
  \BibitemOpen
  \bibfield  {author} {\bibinfo {author} {\bibfnamefont {J.}~\bibnamefont
  {{Stadler}}}, \bibinfo {author} {\bibfnamefont {R.}~\bibnamefont
  {{Mikulla}}}, \ and\ \bibinfo {author} {\bibfnamefont {H.-R.}\ \bibnamefont
  {{Trebin}}},\ }\href@noop {} {\bibfield  {journal} {\bibinfo  {journal}
  {Int.\ J.\ Mod.\ Phys.~C}\ }\textbf {\bibinfo {volume} {8}},\ \bibinfo
  {pages} {1131} (\bibinfo {year} {1997})},\ \bibinfo {note} {{\tt
  http://www.itap.physik.uni-stuttgart.de/\~{}imd/}}\BibitemShut {NoStop}%
\bibitem [{\citenamefont {{Rowley}}\ \emph {et~al.}(1998)\citenamefont
  {{Rowley}}, \citenamefont {{Jemmer}}, \citenamefont {{Wilson}},\ and\
  \citenamefont {{Madden}}}]{itapdb:Rowley1998}%
  \BibitemOpen
  \bibfield  {author} {\bibinfo {author} {\bibfnamefont {A.~J.}\ \bibnamefont
  {{Rowley}}}, \bibinfo {author} {\bibfnamefont {P.}~\bibnamefont {{Jemmer}}},
  \bibinfo {author} {\bibfnamefont {M.}~\bibnamefont {{Wilson}}}, \ and\
  \bibinfo {author} {\bibfnamefont {P.~A.}\ \bibnamefont {{Madden}}},\
  }\href@noop {} {\bibfield  {journal} {\bibinfo  {journal} {J.~Chem.\ Phys.}\
  }\textbf {\bibinfo {volume} {108}},\ \bibinfo {pages} {10209} (\bibinfo
  {year} {1998})}\BibitemShut {NoStop}%
\bibitem [{\citenamefont {{Fincham}}(1994)}]{itapdb:Fincham1994}%
  \BibitemOpen
  \bibfield  {author} {\bibinfo {author} {\bibfnamefont {D.}~\bibnamefont
  {{Fincham}}},\ }\href@noop {} {\bibfield  {journal} {\bibinfo  {journal}
  {Mol.\ Sim.}\ }\textbf {\bibinfo {volume} {13}},\ \bibinfo {pages} {1}
  (\bibinfo {year} {1994})}\BibitemShut {NoStop}%
\bibitem [{\citenamefont {{Brommer}}\ \emph {et~al.}(2010)\citenamefont
  {{Brommer}}, \citenamefont {{Beck}}, \citenamefont {{Chatzopoulos}},
  \citenamefont {{Gähler}}, \citenamefont {{Roth}},\ and\ \citenamefont
  {{Trebin}}}]{Brommer2010}%
  \BibitemOpen
  \bibfield  {author} {\bibinfo {author} {\bibfnamefont {P.}~\bibnamefont
  {{Brommer}}}, \bibinfo {author} {\bibfnamefont {P.}~\bibnamefont {{Beck}}},
  \bibinfo {author} {\bibfnamefont {A.}~\bibnamefont {{Chatzopoulos}}},
  \bibinfo {author} {\bibfnamefont {F.}~\bibnamefont {{Gähler}}}, \bibinfo
  {author} {\bibfnamefont {J.}~\bibnamefont {{Roth}}}, \ and\ \bibinfo {author}
  {\bibfnamefont {H.-R.}\ \bibnamefont {{Trebin}}},\ }\href@noop {} {\bibfield
  {journal} {\bibinfo  {journal} {J.~Chem.\ Phys.}\ }\textbf {\bibinfo {volume}
  {132}},\ \bibinfo {pages} {194109} (\bibinfo {year} {2010})}\BibitemShut
  {NoStop}%
\bibitem [{\citenamefont {{Kresse}}\ and\ \citenamefont
  {{Hafner}}(1993)}]{Kresse1993}%
  \BibitemOpen
  \bibfield  {author} {\bibinfo {author} {\bibfnamefont {G.}~\bibnamefont
  {{Kresse}}}\ and\ \bibinfo {author} {\bibfnamefont {J.}~\bibnamefont
  {{Hafner}}},\ }\href@noop {} {\bibfield  {journal} {\bibinfo  {journal}
  {Phys.\ Rev.~B}\ }\textbf {\bibinfo {volume} {47}},\ \bibinfo {pages} {558}
  (\bibinfo {year} {1993})}\BibitemShut {NoStop}%
\bibitem [{\citenamefont {{Kresse}}\ and\ \citenamefont
  {{Furthmüller}}(1996)}]{Kresse1996}%
  \BibitemOpen
  \bibfield  {author} {\bibinfo {author} {\bibfnamefont {G.}~\bibnamefont
  {{Kresse}}}\ and\ \bibinfo {author} {\bibfnamefont {J.}~\bibnamefont
  {{Furthmüller}}},\ }\href@noop {} {\bibfield  {journal} {\bibinfo  {journal}
  {Phys.\ Rev.~B}\ }\textbf {\bibinfo {volume} {54}},\ \bibinfo {pages} {11169}
  (\bibinfo {year} {1996})}\BibitemShut {NoStop}%
\bibitem [{\citenamefont {{Corona}}\ \emph {et~al.}(1987)\citenamefont
  {{Corona}}, \citenamefont {{Marchesi}}, \citenamefont {{Martini}},\ and\
  \citenamefont {{Ridella}}}]{itapdb:Corona1987}%
  \BibitemOpen
  \bibfield  {author} {\bibinfo {author} {\bibfnamefont {A.}~\bibnamefont
  {{Corona}}}, \bibinfo {author} {\bibfnamefont {M.}~\bibnamefont
  {{Marchesi}}}, \bibinfo {author} {\bibfnamefont {C.}~\bibnamefont
  {{Martini}}}, \ and\ \bibinfo {author} {\bibfnamefont {S.}~\bibnamefont
  {{Ridella}}},\ }\href@noop {} {\bibfield  {journal} {\bibinfo  {journal} {ACM
  Trans. Math. Softw.}\ }\textbf {\bibinfo {volume} {13}},\ \bibinfo {pages}
  {262} (\bibinfo {year} {1987})}\BibitemShut {NoStop}%
\bibitem [{\citenamefont {{Powell}}(1965)}]{itapdb:Powell1965}%
  \BibitemOpen
  \bibfield  {author} {\bibinfo {author} {\bibfnamefont {M.~J.~D.}\
  \bibnamefont {{Powell}}},\ }\href@noop {} {\bibfield  {journal} {\bibinfo
  {journal} {Comp.\ J.}\ }\textbf {\bibinfo {volume} {7}},\ \bibinfo {pages}
  {303} (\bibinfo {year} {1965})}\BibitemShut {NoStop}%
\bibitem [{\citenamefont {{Kresse}}\ and\ \citenamefont
  {{Joubert}}(1999)}]{itapdb:Kresse1999}%
  \BibitemOpen
  \bibfield  {author} {\bibinfo {author} {\bibfnamefont {G.}~\bibnamefont
  {{Kresse}}}\ and\ \bibinfo {author} {\bibfnamefont {D.}~\bibnamefont
  {{Joubert}}},\ }\href@noop {} {\bibfield  {journal} {\bibinfo  {journal}
  {Phys.\ Rev.~B}\ }\textbf {\bibinfo {volume} {59}},\ \bibinfo {pages} {1758}
  (\bibinfo {year} {1999})}\BibitemShut {NoStop}%
\bibitem [{\citenamefont {{Perdew}}\ \emph {et~al.}(1992)\citenamefont
  {{Perdew}}, \citenamefont {{Chevary}}, \citenamefont {{Vosko}}, \citenamefont
  {{Jackson}}, \citenamefont {{Pederson}}, \citenamefont {{Singh}},\ and\
  \citenamefont {{Fiolhais}}}]{pw91}%
  \BibitemOpen
  \bibfield  {author} {\bibinfo {author} {\bibfnamefont {J.~P.}\ \bibnamefont
  {{Perdew}}}, \bibinfo {author} {\bibfnamefont {J.~A.}\ \bibnamefont
  {{Chevary}}}, \bibinfo {author} {\bibfnamefont {S.~H.}\ \bibnamefont
  {{Vosko}}}, \bibinfo {author} {\bibfnamefont {K.~A.}\ \bibnamefont
  {{Jackson}}}, \bibinfo {author} {\bibfnamefont {M.~R.}\ \bibnamefont
  {{Pederson}}}, \bibinfo {author} {\bibfnamefont {D.~J.}\ \bibnamefont
  {{Singh}}}, \ and\ \bibinfo {author} {\bibfnamefont {C.}~\bibnamefont
  {{Fiolhais}}},\ }\href@noop {} {\bibfield  {journal} {\bibinfo  {journal}
  {Phys. Rev. B}\ }\textbf {\bibinfo {volume} {46}},\ \bibinfo {pages} {6671}
  (\bibinfo {year} {1992})}\BibitemShut {NoStop}%
\bibitem [{\citenamefont {{Villars}}\ and\ \citenamefont
  {{Calvert}}(1991)}]{pear}%
  \BibitemOpen
  \bibfield  {author} {\bibinfo {author} {\bibfnamefont {P.}~\bibnamefont
  {{Villars}}}\ and\ \bibinfo {author} {\bibfnamefont {L.~D.}\ \bibnamefont
  {{Calvert}}},\ }\href@noop {} {\emph {\bibinfo {title} {Pearson's Handbook of
  Crystallographic Data for Intermetallic Phases, 2nd Edition, Vol. I}}}\
  (\bibinfo  {publisher} {ASM International, Materials Park, Ohio},\ \bibinfo
  {year} {1991})\ p.\ \bibinfo {pages} {970}\BibitemShut {NoStop}%
\bibitem [{\citenamefont {Weast}(1983)}]{cohes}%
  \BibitemOpen
  \bibinfo {editor} {\bibfnamefont {R.~C.}\ \bibnamefont {Weast}},\ ed.,\
  \href@noop {} {\emph {\bibinfo {title} {CRC Handbook of Chemistry and
  Physics}}}\ (\bibinfo  {publisher} {CRC Press, Boca Raton, FL},\ \bibinfo
  {year} {1983})\BibitemShut {NoStop}%
\bibitem [{\citenamefont {{Siegel}}, \citenamefont {{Hector Jr.}},\ and\
  \citenamefont {{Adams}}(2003)}]{siegel}%
  \BibitemOpen
  \bibfield  {author} {\bibinfo {author} {\bibfnamefont {D.~J.}\ \bibnamefont
  {{Siegel}}}, \bibinfo {author} {\bibfnamefont {L.~G.}\ \bibnamefont {{Hector
  Jr.}}}, \ and\ \bibinfo {author} {\bibfnamefont {J.~B.}\ \bibnamefont
  {{Adams}}},\ }\href@noop {} {\bibfield  {journal} {\bibinfo  {journal}
  {Phys.\ Rev.~B}\ }\textbf {\bibinfo {volume} {67}},\ \bibinfo {pages}
  {092105} (\bibinfo {year} {2003})}\BibitemShut {NoStop}%
\bibitem [{\citenamefont {{Sun}}, \citenamefont {{Stirner}},\ and\
  \citenamefont {{Matthews}}(2006)}]{sun}%
  \BibitemOpen
  \bibfield  {author} {\bibinfo {author} {\bibfnamefont {J.}~\bibnamefont
  {{Sun}}}, \bibinfo {author} {\bibfnamefont {T.}~\bibnamefont {{Stirner}}}, \
  and\ \bibinfo {author} {\bibfnamefont {A.}~\bibnamefont {{Matthews}}},\
  }\href@noop {} {\bibfield  {journal} {\bibinfo  {journal} {Surf. Coat.
  Tech.}\ }\textbf {\bibinfo {volume} {201}},\ \bibinfo {pages} {4205}
  (\bibinfo {year} {2006})}\BibitemShut {NoStop}%
\bibitem [{\citenamefont {{Mackrodt}}\ \emph {et~al.}(1987)\citenamefont
  {{Mackrodt}}, \citenamefont {{Davey}}, \citenamefont {{Black}},\ and\
  \citenamefont {{Docherty}}}]{mack}%
  \BibitemOpen
  \bibfield  {author} {\bibinfo {author} {\bibfnamefont {W.~C.}\ \bibnamefont
  {{Mackrodt}}}, \bibinfo {author} {\bibfnamefont {R.~J.}\ \bibnamefont
  {{Davey}}}, \bibinfo {author} {\bibfnamefont {S.~N.}\ \bibnamefont
  {{Black}}}, \ and\ \bibinfo {author} {\bibfnamefont {R.}~\bibnamefont
  {{Docherty}}},\ }\href@noop {} {\bibfield  {journal} {\bibinfo  {journal} {J.
  Cryst. Growth}\ }\textbf {\bibinfo {volume} {80}},\ \bibinfo {pages} {441}
  (\bibinfo {year} {1987})}\BibitemShut {NoStop}%
\bibitem [{\citenamefont {{Manassidis}}, \citenamefont {{De Vita}},\ and\
  \citenamefont {{Gillan}}(1993)}]{mana}%
  \BibitemOpen
  \bibfield  {author} {\bibinfo {author} {\bibfnamefont {I.}~\bibnamefont
  {{Manassidis}}}, \bibinfo {author} {\bibfnamefont {A.}~\bibnamefont {{De
  Vita}}}, \ and\ \bibinfo {author} {\bibfnamefont {M.~J.}\ \bibnamefont
  {{Gillan}}},\ }\href@noop {} {\bibfield  {journal} {\bibinfo  {journal}
  {Surf. Sci. Lett.}\ }\textbf {\bibinfo {volume} {285}},\ \bibinfo {pages}
  {L517} (\bibinfo {year} {1993})}\BibitemShut {NoStop}%
\bibitem [{\citenamefont {{Pinto}}, \citenamefont {{Nieminen}},\ and\
  \citenamefont {{Elliot}}(2004)}]{pinto}%
  \BibitemOpen
  \bibfield  {author} {\bibinfo {author} {\bibfnamefont {H.~P.}\ \bibnamefont
  {{Pinto}}}, \bibinfo {author} {\bibfnamefont {R.~M.}\ \bibnamefont
  {{Nieminen}}}, \ and\ \bibinfo {author} {\bibfnamefont {S.~D.}\ \bibnamefont
  {{Elliot}}},\ }\href@noop {} {\bibfield  {journal} {\bibinfo  {journal}
  {Phys.\ Rev.~B}\ }\textbf {\bibinfo {volume} {70}},\ \bibinfo {pages}
  {125402} (\bibinfo {year} {2004})}\BibitemShut {NoStop}%
\bibitem [{\citenamefont {{Rog}}\ \emph {et~al.}(2003)\citenamefont {{Rog}},
  \citenamefont {{Murzyn}}, \citenamefont {{Hinsen}},\ and\ \citenamefont
  {{Kneller}}}]{Rog2003}%
  \BibitemOpen
  \bibfield  {author} {\bibinfo {author} {\bibfnamefont {T.}~\bibnamefont
  {{Rog}}}, \bibinfo {author} {\bibfnamefont {K.}~\bibnamefont {{Murzyn}}},
  \bibinfo {author} {\bibfnamefont {K.}~\bibnamefont {{Hinsen}}}, \ and\
  \bibinfo {author} {\bibfnamefont {G.~R.}\ \bibnamefont {{Kneller}}},\
  }\href@noop {} {\bibfield  {journal} {\bibinfo  {journal} {J.~Comput.\
  Chem.}\ }\textbf {\bibinfo {volume} {24}},\ \bibinfo {pages} {657} (\bibinfo
  {year} {2003})}\BibitemShut {NoStop}%
\bibitem [{\citenamefont {{Lodziana}}\ and\ \citenamefont
  {{Parlinski}}(2003)}]{Lodziana2003}%
  \BibitemOpen
  \bibfield  {author} {\bibinfo {author} {\bibfnamefont {Z.}~\bibnamefont
  {{Lodziana}}}\ and\ \bibinfo {author} {\bibfnamefont {K.}~\bibnamefont
  {{Parlinski}}},\ }\href@noop {} {\bibfield  {journal} {\bibinfo  {journal}
  {Phys.\ Rev.~B}\ }\textbf {\bibinfo {volume} {67}},\ \bibinfo {pages}
  {174106} (\bibinfo {year} {2003})}\BibitemShut {NoStop}%
\bibitem [{\citenamefont {{Morrissey}}\ and\ \citenamefont
  {{Carter}}(1984)}]{morr}%
  \BibitemOpen
  \bibfield  {author} {\bibinfo {author} {\bibfnamefont {K.~J.}\ \bibnamefont
  {{Morrissey}}}\ and\ \bibinfo {author} {\bibfnamefont {C.~B.}\ \bibnamefont
  {{Carter}}},\ }\href@noop {} {\bibfield  {journal} {\bibinfo  {journal} {J.
  Am. Ceram. Soc.}\ }\textbf {\bibinfo {volume} {67}},\ \bibinfo {pages} {292}
  (\bibinfo {year} {1984})}\BibitemShut {NoStop}%
\bibitem [{\citenamefont {{Thomson}}, \citenamefont {{Hsieh}},\ and\
  \citenamefont {{Rana}}(1970)}]{lattice}%
  \BibitemOpen
  \bibfield  {author} {\bibinfo {author} {\bibfnamefont {R.}~\bibnamefont
  {{Thomson}}}, \bibinfo {author} {\bibfnamefont {C.}~\bibnamefont {{Hsieh}}},
  \ and\ \bibinfo {author} {\bibfnamefont {V.}~\bibnamefont {{Rana}}},\
  }\href@noop {} {\bibfield  {journal} {\bibinfo  {journal} {J. Appl. Phys.}\
  }\textbf {\bibinfo {volume} {42}},\ \bibinfo {pages} {3154} (\bibinfo {year}
  {1970})}\BibitemShut {NoStop}%
\bibitem [{MegaMol()}]{megamol}%
  \BibitemOpen
  MegaMol,\ \href {http://www.visus.uni-stuttgart.de/megamol} {}\bibinfo {note}
  {{\url{http://www.visus.uni-stuttgart.de/megamol/}}}\BibitemShut {NoStop}%
\end{thebibliography}
\end{document}